\documentstyle[preprint,aps,amssymb]{revtex}
\begin{document}
\draft
\preprint{RUB-TPII-36/93 \\}
\date{February, 1993}
\title{
HETEROTIC APPROACH TO THE NUCLEON DISTRIBUTION AMPLITUDE}
\author{M. BERGMANN\footnote{E-mail:
michaelb@hadron.tp2.ruhr-uni-bochum.de}
and N. G. STEFANIS\thanks{Work supported in part by the
Deutsche Forschungsgemeinschaft}}

\address{Institut f\"ur Theoretische Physik II  \\
Ruhr-Universit\"at Bochum  \\
D-44780 Bochum, Germany}

\maketitle

\begin{abstract}
We give an in-depth analysis of the determination procedure
of the recently proposed heterotic nucleon distribution
amplitude which hybridizes the best features of the
Chernyak-Ogloblin-Zhitnitsky and the Gari-Stefanis models.
With respect to the QCD sum-rule constraints, optimized
versions of these amplitudes are derived in terms of which
a "hybridity" angle can be introduced to systematically
classify all models.
\end{abstract}
Large-momentum transfer exclusive reactions comprise two
mechanisms based on the factorization theorem~\cite{LB80,Mue81}:
the hard-gluon exchanges that describe short-range interactions via
perturbation theory and the long-range (nonperturbative) soft processes
that ensure confinement in the hadronic states. Thus, such processes
depend in detail on the composition of the hadron wave function or
more precisely, the hadron distribution amplitude:
$\Phi (x_{i},Q^{2})$.
In a physical gauge as the light-cone gauge
$A^{+}=A^{0}+A^{3}=0$,
this is the fundamental covariant amplitude describing the correlations
between massless valence quarks with fractional momenta
$0 \le x_{i} = \frac{k_{i}^{+}}{p^{+}} \le 1$
($\sum_{i}^{}x_{i} =1$)
and relative transverse momenta
${\bf k}_{\bot}^{(i)}$, $p^{+}$
being the longitudinal momentum of the incoming hadron. Then,
apart from logarithmic scaling violations,
\begin{equation}
   \Phi (x_{i},Q^{2}) = \int_{}^{Q^{2}}[d^{2}k_{\bot}]\,
   \psi (x_{i},{\bf k}_{\bot}^{(i)}),
\end{equation}
where
\hbox{$[d^{2}k_{\bot}]=16\pi ^{3}\delta ^{(2)}(\sum_{i=1}^{3}\,
{\bf k}_{\bot}^{(i)})\prod_{i=1}^{3}\,\left[\frac{d^{2}k_{\bot}^{(i)}}
{16\pi ^{3}}\right]$}
and
$\psi (x_{i},{\bf k}_{\bot}^{(i)})$
is the lowest-twist Fock-space projection amplitude of the hadron bound
state. The actual calculation of $\Phi (x_{i},Q^{2})$ from QCD requires
nonperturbative methods such as QCD sum rules~\cite{CZ84a}, lattice
gauge theory~\cite{KP85,RSS87} or the direct diagonalization of the
light-cone hamiltonian within a discretized light-cone setup (see, eg.,
\cite{BP91}).

In the context of QCD sum rules, useful theoretical constraints on the
nucleon distribution amplitude have been derived by a number of
authors~\cite{CZ84b,KS87,COZ89a}. They have been used to obtain model
distribution amplitudes for the
nucleon~\cite{CZ84b,KS87,COZ89a,GS86}
and the $\Delta ^{+}(1232)$ isobar~\cite{CP88,FZOZ88,SB93a}
in terms of the eigenfunctions of the evolution equation~\cite{LB80}
represented by Appell polynomials~\cite{Erd53}. However, the
inevitable model purposes lead to emphasis on different facets of the
complete picture. As a result, the Chernyak-Ogloblin-Zhitnitsky (COZ)
model~\cite{COZ89a} predicts $R\equiv \vert G_{M}^{n} \vert /
G_{M}^{p} \approx 0.5$, whereas the Gari-Stefanis (GS)
model~\cite{GS86}, inspired by a semiphenomenological
analysis~\cite{GK85}, gives by construction a small $R$, {\it viz.},
$R=0.097$.

One would like the sum rules eventually to force the choice between
COZ-type and GS-type models on grounds of mathematical, rather than
phenomenological, necessity. In a recent publication~\cite{SB93b}
(see also~\cite{Ste92}) we have suggested that there is a third {\it
distinct} possibility for the nucleon distribution amplitude, which
actually hybridizes features of both types of models into a single mold.
What is novel is the requirement that these models should be regarded as
different aspects of a more fundamental unifying structure, we termed
the "heterotic" solution. In the present work our interest in the model
is primarily theoretic. We shall show that the heterotic solution is
uniquely determined in the parameter space spanned by the expansion
coefficients on the orthonormalized eigenfunctions of the evolution
equation. With respect to a $\chi ^{2}$ criterion to describe the
deviations from the sum rules, it corresponds to that local minimum
which is associated with the smallest possible value of $R$, notably
$0.1$, compatible with the sum-rule constraints~\cite{COZ89a,KS87}.

In~\cite{SB93b,Ste92} we have given evidence that the heterotic
model is the single best way to promote agreement between theory and the
data on a variety of exclusive processes including elastic
electron-nucleon scattering and charmonium decays into $p \bar p$.

In searching for a starting point from which to develop a credible
nucleon distribution amplitude, we employ the ansatz described
in~\cite{Ste89}. Then the mixed-symmetry distribution amplitude
$\Phi _{N}(x_{i})=V(x_{i})-A(x_{i})$~\cite{HKM75} at fixed scale
$Q^{2}$ can be reconstructed from its moments
\begin{equation}
  \Phi _{N}^{(n_{1}n_{2}n_{3})} =
  \int_{0}^{1}[dx]\, x_{1}^{n_{1}}x_{2}^{n_{2}}x_{3}^{n_{3}}\,
  \Phi _{N}(x_{i}).
\end{equation}
To this end, consider
\begin{equation}
  (z\cdot p)^{-(n_{1}+n_{2}+n_{3})}
  \prod_{i=1}^{3}\left(iz\cdot
  \frac{\partial}{\partial z_{i}}\right)^{n_{i}}
  \Phi _{N}(z_{i}\cdot p)\big\vert _{z_{i}=0} =
  \Phi _{N}^{(n_{1}n_{2}n_{3})},
\end{equation}
where $z$ is an auxiliary lightlike vector ($z^{2}=0$).
To determine the moments, a short-distance operator-product expansion is
performed at some spacelike momentum where quark-hadron duality is
supposed to be valid. Such a computation involves correlation functions
of the form ($-q^{2}=Q^{2}$)
\begin{eqnarray}
  I^{\,(n_{1}n_{2}n_{3},m)}(q,z)
& = &
  i\int_{}^{}
  d^{4}x \, e^{iq\cdot x}
  <\Omega\vert T\bigl (O_{\gamma}^{\,(n_{1}n_{2}n_{3})}(0)
  \hat O_{\gamma\prime}^{\,(m)}(x)\bigr )\vert\Omega >(z\cdot \gamma )_
  {\gamma \gamma\prime}
\nonumber \\
& = &
  (z\cdot q)^{\,{n_{1}+n_{2}+n_{3}+m+3}}I^{\,(n_{1}n_{2}n_{3},m)}
  (q^{2}),
\end{eqnarray}
where $O_{\gamma}^{\,(n_{1}n_{2}n_{3})}$ are
appropriate~\cite{COZ89a,KS87} three-quark operators interpolating
between the proton state and the vacuum. Since they contain derivatives,
their matrix elements are related to moments of the covariant amplitudes
$V$, $A$, and $T$:
\begin{equation}
  <\Omega \vert O_{\gamma}^{(n_{1}n_{2}n_{3})}(0)\vert P(p)> =
  f_{N}(z\cdot p)^{\,{n_{1}+n_{2}+n_{3}+1}}N_{\gamma}\,
  O^{\,(n_{1}n_{2}n_{3})}.
\end{equation}
The factor $(z\cdot \gamma )_{\gamma\gamma\prime}$ serves to
separate out the leading twist structure in the correlator; $N_{\gamma}$
is the proton spinor, and $f_{N}$ denotes the "proton decay constant".

Because $\Phi _{N}(x_{i})$ must be a solution of the evolution equation,
it can be expressed in the form
\begin{equation}
  \Phi _{N}(x_{i}, Q^{2})=\Phi _{as}(x_{i})\sum_{n=0}^{\infty}B_{n}
     \tilde \Phi _{n}(x_{i})\biggl (
     \frac{\alpha _{s}(Q^{2})}{\alpha _{s}(\mu ^{2})}\biggr )
     ^{\gamma _{n}},
\end{equation}
in which $\{\Phi _{n}\}_{0}^{\infty}$ are the eigenfunctions of the
interaction kernel of the evolution equation, orthonormalized within
a basis of Appell polynomials of degree $M$, and
$\Phi _{as}(x_{i})=120x_{1}x_{2}x_{3}$ (see~\cite{LB80}).
[The conventions and analytical expressions given in~\cite{Ste89} are
used.] The corresponding eigenvalues $\gamma _{n}$ are identical
with the anomalous dimensions of multiplicatively renormalizable baryonic
operators of twist three~\cite{Pes79}. Because the $\gamma _{n}$ are
positive fractional numbers increasing with $n$, higher terms in the
expansion (6) are gradually suppressed.

The expansion coefficients $B_{n}$ can be determined by means of moments
inversion on account of the orhogonality of the eigenfunctions
$\{\tilde \Phi _{n}(x_{i})\}$:
\begin{equation}
  B_{n}(\mu ^{2})=\frac{N_{n}}{120}\int_{0}^{1}[dx]\, \tilde
  \Phi _{n}(x_{i})\Phi _{N}(x_{i}, \mu ^{2}).
\end{equation}
Evaluating the correlator in Eq.~(4) for $n_{1}+n_{2}+n_{3}\le 3$ and
$m=1$, the coefficients $B_{n}$ for $n=0,1\ldots ,5$ (i.e., $max(M)=2$)
can be computed upon imposing on the moments of $\Phi _{N}$ the sum-rule
constraints estimated by COZ in~\cite{COZ89a}.

The starting point of our method is to consider the hyperspace
induced by the expansion coefficients $B_{n}$ and look for solutions of
the form given by Eq.~(6) complying with the COZ sum-rule requirements.
The deviations from the sum rules are parametrized by a $\chi ^{2}$
criterion which accounts for the order of the moments (corresponding to
the superior stability of lower moments relatively to the higher
ones~\cite{Ste89}). For every moment $m_{k}$ $(k=1,\ldots ,18)$, we
define
\begin{equation}
  \chi ^{2}_{k}=(\chi ^{2}_{k,(a)}+\chi ^{2}_{k,(b)})
  \,[1\, - \,\Theta (m_{k}-M_{k}^{min})\Theta (M_{k}^{max}-m_{k})]
\end{equation}
with
\begin{equation}
  \chi ^{2}_{k,(a)}=min ( \vert M_{k}^{min}-m_{k}\vert,
  \vert m_{k}-M_{k}^{max}\vert )N_{k}^{-1},
\end{equation}
where
$
 N_{k}=\vert M_{k}^{min} \vert
$
or
$
 \vert M_{k}^{max} \vert
$,
whether $m_{k}$ lies on the left or on the right hand side of the
corresponding sum-rule interval
($\chi ^{2}_{\text{tot}}=\sum_{k}^{}\chi ^{2}_{k}$).

Deviations from lower-order moments are weighted by a larger
penalty factor than those from higher-order moments:
\begin{equation}
  \chi ^{2}_{k,(b)}=\left\{
                           \begin{array}{ll}
                           100, &   1\leq k\leq  3 \\
                           10,  &   4\leq k\leq  9 \\
                            1,  &  10\leq k\leq 18.
                           \end{array}
                    \right.
\end{equation}
In this way we were able to filter out a COZ-type solution (designated
as $COZ^{\text{opt}}$) which is the absolute minimum of $\chi ^{2}$ and
thus provides the best possible agreement with the sum rules. Evolving
that solution towards lower values of $R$, we generated a series of
local minima of $\chi ^{2}$, plotted in Fig.~1(a) in the ($B_{4},R$)
plane. [The assignments of models to symbols are given in Table~1.]
The dashed line is a convenient fit expressed by
$
 R = 0.436415 - 0.005374 B_{4} - 0.000197 B_{4}^{2}
$.
The ultimate local minimum accessible from $COZ^{\text{opt}}$ on that
$\chi ^{2}$ orbit corresponds to $R=0.1$---the heterotic solution.
This solution, although degenerated with respect to $R$, is {\it
distinct} from the GS one. The latter, as well as its optimized versions
we determined, constitute an isolated region (an "island") in the
parameter space. Technically this means that they correspond to local
minima of $\chi ^{2}$ at considerably lower levels of accuracy so that
they are separated from the COZ-Het orbit by a large $\chi ^{2}$
barrier.
Without considering here applications to physical processes, we only
mention that the predictions extracted from the heterotic model are
dramatically different compared to those following from the GS
model~\cite{SB93b,Ste92,BS93a}.
Also, the CZ and KS models, although in the vicinity of the COZ-Het orbit,
are actually isolated points because they correspond to much
larger $\chi ^{2}$ values (see Fig.~1(b) in correspondence with
Table~1).
One has to exercise a certain amount of care to be sure that these are
the only regions in the parameter space contributing to the sum rules at
the desired level of accuracy.

The general picture which emerges is a pattern of nucleon distribution
amplitudes which develops into several regions of different $\chi ^{2}$
dependence (Fig.~1(b)). The new heterotic amplitude and the original COZ
one correspond to similar $\chi ^{2}$ values. In contrast, the previous
models (CZ, GS, and KS) are characterized by large deviations from the
sum rule requirements. Optimum consistency with the sum rules is provided
by the amplitude $COZ^{\text{opt}}$, but this amplitude is not favored
because it fails to predict hadronic observables in agreement with the
existing data~\cite{BS93a}. It is remarkable that the heterotic
solution~\cite{SB93b} matches the King-Sachrajda~\cite{KS87} sum-rule
constraints better than the original COZ amplitude. This is particularly
important because the KS results have been independently verified
in~\cite{CP88}.

The variation with shape of the nucleon distribution amplitude with $R$
is shown graphically, obtained by interpolating between the calculated
version of the optimized COZ amplitude and the heterotic one, in
Fig~2. Note that although the heterotic solution was {\it continuously}
evolved from $COZ^{\text{opt}}$, its geometrical characteristics
are something of a hybrid between COZ-type and GS-type amplitudes.
[The profile of $\Phi _{N}^{\text{Het}}$ is shown in~\cite{SB93a,Ste92}.]
This heterotic character becomes apparent by considering the
corresponding covariant distribution amplitudes $V$, $A$, and $T$
\cite{HKM75}, shown in Fig.~3. While $V_{\text{Het}}$ has a symmetry
pattern similar to that of $V_{\text{GS}}$ (one main maximium),
$T_{\text{Het}}$ is characterized by two maxima, as $T_{\text{COZ}}$.
The inverse heterotic combination does not belong to the COZ-Het $\chi
^{2}$-orbit. This type of solution can be realized by the following
expansion coefficients: $B_{1}=4.3025$, $B_{2}=1.5920$, $B_{3}=1.9675$,
$B_{4}=-19.6580$, and $B_{5}=3.3531$, corresponding to
$\vartheta=24.44^{\circ}$ (see Eq.~(11)) and $\chi ^{2}=30.634$.
Remarkably, this solution yields $R=0.448$, i.e., it is COZ-like
(albeit in a different region of the parameter space).
The dynamic implications of our solutions for various physical
observables will be discussed elsewhere.

To put these remarks on mathematical grounds, we propose to consider
a classification scheme of the various nucleon distribution amplitudes
based on the observation that the optimized versions of the COZ-type
and GS-type amplitudes we derived are almost orthogonal to each other
with respect to the weight $w(x_{i})=\Phi _{as}(x_{i})/120$
\cite{Erd53}.
Their normalized inner product $(COZ^{\text{opt}},GS^{\text{opt}})$
yields $0.1607$, which corresponds to an angle of $80.8^{\circ}$. Thus
they form a quasi-orthogonal basis and can be used to classify the
nucleon distribution amplitudes in terms of a "hybridity" angle
$\vartheta$, defined by
\begin{equation}
  \vartheta = \arctan\left(\frac{\kappa _{1}(i)}
{\kappa_{2}(i)}\right),
\end{equation}
where $\kappa _{1}(i)=(GS^{\text{opt}},i)$ and
$\kappa _{2}(i)=(COZ^{\text{opt}},i)$
and the index $i$ denotes one of the nucleon distribution amplitudes
listed in Table~1. The hybridity angle parametrizes the mingling
of geometrical characteristics attributed to COZ-like and GS-like
amplitudes and provides a quantitative measure for their presence in
any solution conforming with the sum-rule constraints (Fig.~3(c)).
The superimposed dashed line is a fit given by
$
 \vartheta /[deg] = 8.5693 + 0.0160B_{4} + 0.0073B_{4}^{2}-
 0.00067B_{4}^{3}
$.
The mixing of different geometrical characteristics is particularly
relevant for the heterotic solution, which generically
amalgamates features of both types of amplitudes. In this role, the
heterotic model has the special virtue of simultaneously fitting the
twin hopes for making reliable predictions with respect to the
experimental data while being in agreement with the sum-rule
constraints~\cite{SB93a,SB93b,BS93a}. The other considered models
can be tuned to fit some aspects of the data, but never all aspects
simultaneously.

  From the above considerations we conclude that a key clue to the
determination of nucleon distribution amplitudes on the basis of QCD
sum rules is the resulting pattern of order effected in Fig.~1,
the ordering parameters being the expansion coefficient $B_{4}$
and the hybridity angle $\vartheta$. This classification scheme can
valuably supplement the standard fitting procedure to the sum rules when
enlarging the basis of eigenfunctions of the evolution kernel to
include higher-order Appell polynomials. On the phenomenological side,
fixing the value of the ratio $R$ by experiment, one could use Fig.~1(a)
to extract the corresponding nucleon distribution amplitude consistent
with the sum-rule constraints.

\begin{table}
\caption{Expansion coefficients $B_{n}$, hybridity angle
$\vartheta$, and values of the ratio
$R=\vert G_{M}^{n} \vert / G_{M}^{p}$,
in correspondence with those for $\chi ^{2}$
for the various nucleon distribution amplitudes discussed in
the text.}
\begin{tabular}{lrrrrrrrrc}
 Model      & $B_{1}$\ \ &  $B_{2}$\ \ &   $B_{3}$\ \  &    $B_{4}$\ \ &
$B_{5}$\ \ &
$\vartheta$[deg]\ \  &  R\ \ & $\chi^{2}$\ & Symbol \cr
\hline
 $Het.     $ & 3.4437 &  1.5710 &   4.5937  &   29.3125 &    -0.1250  &
-1.89  &    .104 &  33.48 &{\Large ${\bullet}$} \cr
 $COZ^{opt}$ & 3.5268 &  1.4000 &   2.8736  &   -4.5227 &     0.8002  &
9.13  &    .465 &   4.49 &$\blacksquare$    \cr
 $GS^{opt} $ & 3.9501 &  1.5273 &  -4.8174  &    3.4435 &     8.7534  &
80.87  &    .095 &  54.95 &$\blacktriangle$ \cr
 $GS^{min} $ & 3.9258 &  1.4598 &  -4.6816  &    1.1898 &     8.0123  &
80.19  &    .035 &  54.11 &$\blacktriangledown$ \cr
 $CZ       $ & 4.3050 &  1.9250 &   2.2470  &   -3.4650 &     0.0180  &
13.40  &    .487 & 250.07 &$\blacklozenge$ \cr
 $COZ      $ & 3.6750 &  1.4840 &   2.8980  &   -6.6150 &     1.0260  &
10.16  &    .474 &  24.64 &$\Box$   \cr
 $KS       $ & 3.2550 &  1.2950 &   3.9690  &    0.9450 &     1.0260  &
2.47  &    .412 & 116.35 &$\Diamond$ \cr
 $GS       $ & 4.1045 &  2.0605 &  -4.7173  &    5.0202 &     9.3014  &
78.87  &    .097 & 270.82 &$\bigtriangleup$ \cr
\hline
Samples      & \multicolumn{5}{c}{\quad} &             &          &  &
\cr
\hline
 $   0     $ & 3.3125 &  1.4644 &   3.1438  &   -1.0000 &     0.8750  &
7.67  &    .441 &  4.63 &$+$ \cr
 $   1     $ & 3.2651 &  1.4032 &   3.5466  &    2.8685 &     1.7954  &
8.94  &    .405 &  5.11 &$+$ \cr
 $   2     $ & 3.4026 &  1.4917 &   3.0629  &    7.3430 &     0.6719  &
8.75  &    .385 & 16.07 &$+$ \cr
 $   3     $ & 3.7225 &  1.5030 &   3.6592  &   10.7265 &     1.5154  &
9.29  &    .355 & 17.78 &$+$ \cr
 $   4     $ & 3.8407 &  1.4968 &   3.2142  &   14.4093 &     0.8757  &
10.49  &    .325 & 19.41 &$+$ \cr
 $   5     $ & 3.6544 &  1.4000 &   3.0993  &   15.5614 &    -0.1329  &
6.35  &    .305 & 18.15 &$+$ \cr
 $   6     $ & 3.8607 &  1.4000 &   3.2375  &   19.8571 &    -0.1635  &
6.32  &    .255 & 20.57 &$+$ \cr
 $   7     $ & 3.9783 &  1.4000 &   3.2706  &   22.4194 &    -0.4805  &
5.29  &    .225 & 21.69 &$+$ \cr
 $   8     $ & 4.1547 &  1.4000 &   3.3756  &   26.1305 &    -0.5855  &
5.02  &    .175 & 23.53 &$+$ \cr
 $   9     $ & 3.4044 &  1.5387 &   4.3094  &   25.5625 &     0.0625  &
.01  &    .153 & 30.80 &$+$ \cr
\end{tabular}
\end{table}

\begin{figure}
\caption{Classification scheme of nucleon distribution
         amplitudes conforming with the sum rules (Table~1).
(a) The ratio $R=\vert G_{M}^{n}\vert /G_{M}^{p}$
    as a function of the expansion coefficient $B_{4}$.
    The positions of the $\chi ^{2}$ minima are indicated.
(b) Distribution of local minima of $\chi ^{2}$ (on a logarithmic scale)
    plotted vs. the expansion coefficient $B_{4}$.
(c) Pattern of nucleon distribution amplitudes parametrized by the
    hybridity angle $\vartheta$, defined in Eq.~(11), vs. the expansion
    coefficient $B_{4}$. The dashed curves are the fits described in the
    text.}
\end{figure}

\begin{figure}
\caption{A two-dimensional example showing how the profile of the
nucleon distribution amplitude $\Phi _{N}$ changes along the COZ-Het
$\chi ^{2}$ orbit. The amplitudes are labeled by the corresponding
$R$-value.}
\end{figure}

\begin{figure}
\caption{Covariant distribution amplitudes $V$, $A$, and $T$
associated with the optimized versions of the COZ and GS models in
comparison with the heterotic model.}
\end{figure}
\end{document}